\documentstyle[12pt,a4]{article}
\def\lsim{\:\raisebox{-0.5ex}{$\stackrel{\textstyle<}{\sim}$}\:}
\def\gsim{\:\raisebox{-0.5ex}{$\stackrel{\textstyle>}{\sim}$}\:}
\begin{document}
\begin{flushright}
{\sl TIFR/TH/95-27}
\end{flushright}
\bigskip\bigskip
\begin{center}
{\Large{\bf THE TOP STORY}}\footnote{Based on the talks given at The
Seminar on Recent Advances in Theoretical Physics, Hyderabad, 7-8
April 1995 and The Meeting on Particle Physics, Calcutta, 20-21 April
1995.} \\
\bigskip\bigskip\bigskip
{\large D.P. Roy} \\
\bigskip
Tata Institute of Fundamental Research \\
Homi Bhabha Road, Bombay - 400 005, India \\
\bigskip\bigskip\bigskip
{\large\underbar{Abstract}}
\end{center}
\bigskip

This is an overview of top quark search, with particular emphasis on
the more recent results.  After a brief introduction to the basic
constituents of matter and their interactions, I shall discuss the
indirect evidences for the existence of top quark and its mass from
LEP and finally the direct observation of a top quark signal recently
reported from the Tevatron collider.  I shall try to put these results
in perspective and provide some insight into the physics issues
involved in the top quark search.

\newpage

\section{Basic Constituents of Matter and Their
Interactions}

As per our current wisdom the basic constituents of matter are a dozen of
fermions : the six leptons -- electron, muon, tau and their associated
neutrinos; and the six quarks -- up, down, strange, charm, bottom and top.
They can be arranged as three pairs or generations of leptons and
quarks, which are shown below in increasing order or mass.
\begin{center}
Table 1
\end{center}
\[
\begin{tabular}{|l|c|c||c|c|c|}
\hline
Leptons & $Q$ & $T^3$ & Quarks & $Q$ & $T^3$ \\
\hline
$\nu_e \ \nu_\mu \ \nu_\tau$ & ~0 & ~1/2 & $u \ c \ t$ & ~2/3 & ~1/2 \\
$e \ \ \mu \ \ \tau$ & -1 & -1/2 & $d \ s \ b$ & -1/3 & -1/2 \\
\hline
\end{tabular}
\]
Each pair represents two states differing by 1 unit of
electric charge $Q$, which correspond to the two eigenstates of weak
isospin $T^3 = \pm 1/2$.  In addition, the quarks possess a new type
of charge called the colour charge, which is responsible for their
strong interaction.

Many of these fundamental particles, the $\tau$ lepton and the charm
and bottom quarks, were discovered during the seventies.  Thus by the
end of the seventies all of them had been seen except for the last and
the heaviest one -- i.e. the top quark.  Consequently the top quark
search has been an area of intense activity since the early eighties.
The evidence for top has built up step by step from many indirect
experiments during this period, culminating in the direct observation
of a top quark signal recently reported from the Tevatron collider.  I
shall give an overview of this top search programme, concentrating on
the more recent results.  To facilitate this discussion and fix the
notation let me briefly recall the basic interactions between these
fundamental particles.

Apart from gravitation, which is too weak to be of interest to our
discussion of subatomic particles, there are 3 basic interactions --
strong, electromagnetic and weak.  They are all gauge interactions
mediated by vector particles with couplings proportional to the
corresponding gauge charge.  The strong interaction (QCD) is mediated
by the exchange of massless vector gluons, which couple to all
coloured particles (quarks) with coupling proportional to the colour
charge $C$ (Fig. 1a).  This is analogous to the electromagnetic
interaction (QED), mediated by the exchange of massless vector photon,
which couples to all charged particles (quarks and charged leptons)
with coupling proportional to the electric charge $Q$ (Fig. 1b).  It
is customary to write the strong coupling constant as $$
\alpha_s = g^2_s/4\pi
\eqno (1)
$$
in analogy with the fine structure constant
$$
\alpha = e^2/4\pi.
\eqno (2)
$$
The weak interactions are mediated by the massive charged and neutral
vector bosons $W^\pm$ and $Z^0$.  The charged $W$ boson couples to each of
the above pairs of quarks and leptons with the same universal coupling $g$
(Fig. 1c), where the combination of the Dirac $\gamma$ matrices correspond
to the famous $V-A$ (Lefthanded) form of the charged current weak
interaction corresponding to the gauge group $SU(2)_L$.  The neutral $Z$
boson couples to each quark and lepton (Fig. 1d) with couplings specified
by the standard electro-weak model of Glashow, Weinberg and Salam.  Here
the weak and electromagnetic interactions are unified into a $SU(2)_L
\times U(1)$ gauge interaction, mediated by a charge triplet of gauge
bosons $W^{\pm,0}$ with couplings proportional to the three $SU(2)$
generators $T^{\pm,3}$ (weak isospin) and a charge singlet $B^0$ with
coupling proportional to the $U(1)$ generator (weak hypercharge).  The two
neutral bosons get mixed to give the physical $Z$ boson and photon.  It is
customary to use the $SU(2)$ coupling $g$ and the mixing angle $\theta_W$
as the two independent parameters.  Then the physical $Z$ coupling is
given by [1]
$$
{g \over \cos \theta_W} \left[T^3\gamma^\mu {(1 - \gamma^5) \over 2} -
\sin^2 \theta_W Q \gamma^\mu\right]
\eqno (3)
$$
and the physical photon coupling is related to these parameters by
$$
g = e/\sin \theta_W.
\eqno (4)
$$
The universality of the charged $W$ coupling is of course a simple
reflection of the fact that all the fermions appear in identical
(doublet) representations of the weak isospin group $SU(2)$; i.e. they all
possess the same $SU(2)$ charge.

In the standard model the $SU(2)$ gauge symmetry is spontaneously
broken by the Higgs mechanism to give masses to the $W$ and $Z$ bosons
as well as the fermions.  Since the $W$ and $Z$ bosons acquire their
masses and hence longitudinal components by absorbing the Higgs
scalars, the longitudinal $W$ and $Z$ bosons have Higgs like Yukawa
couplings to the fermions that are proportional to the corresponding
fermion masses.  This is important in the case of top quark due to its
large mass; and we shall see later that it plays a crucial role in the
indirect estimate of the top quark mass.  On the other hand the $W$
and $Z$ boson interactions with all the lighter fermions are
adequately described by the gauge couplings of (3,4), which are
unaffected by the symmetry breaking.  This plays an important role in
extracting indirect evidence for the existence of top quark, as we see
below.

\section{Indirect Evidence for the Existence of Top Quark}

As we see from (3), the $Z$ boson coupling to bottom quark depends
sensitively on its isospin $T^3_b$, which is $-1/2$ or $0$ depending
on whether it is accompanied by a top quark as its isospin partner or
not.  Thus a measurement of the $Z$ boson coupling to the bottom quark
can distinguish between the two alternatives.  There are a number of
indirect evidences for the existence of top quark, based on this
principle.  They come from 1) the forward-backward asymmetry observed
in $e^+e^- \rightarrow \bar b b$ at PETRA energy [2]; 2) the absence
of flavour changing neutral current decay of $b$ quark [3] as
measured at CESR [4]; 3) the absence of large (tree-level) $B_d - \bar
B_d$ mixing [5] as measured at DORIS and CESR [6]; and finally 4) the
direct measurement of the $Z \rightarrow \bar bb$ width at LEP [7].
We shall concentrate on the last process, which provides by far the
cleanest and strongest indirect evidence
for top [8].  A comprehensive account of all the four processes can be
found in [9].

The $Z \rightarrow \bar bb$ decay width has been measured to good
precission at LEP giving [7]
$$
\Gamma (Z \rightarrow \bar bb) = 385 \pm 6 ~{\rm MeV}.
\eqno (5)
$$
 From the $Z\bar bb$ coupling of (3), one can easily estimate this
quantity to be [8]
$$
\Gamma(Z \rightarrow \bar bb) = {G_F M^3_Z \over \sqrt{2} \pi} \left(1
+ {\alpha_s \over \pi}\right) \left[\left(T^3_b - Q_b \sin^2
\theta_W\right)^2 + \left(Q_b \sin^2 \theta_W\right)^2\right]
\eqno (6)
$$
including the small QCD correction.  The effective Fermi coupling [7]
$$
G_F = {\sqrt{2} g^2 \over 8M^2_W} = 1.166 \times 10^{-5} ~{\rm
GeV}^{-2}.
\eqno (7)
$$
Thanks to the small values of $\sin^2 \theta_W$ and the $b$ quark
charge, the width depends crucially on the isospin assignment of $b$.
One gets
$$
\Gamma(Z \rightarrow \bar bb) = 381 ~{\rm MeV} ~(24 ~{\rm MeV})~{\rm
for}~ T^3_b = - {1\over2} (0).
\eqno (8)
$$
Comparing these predictions with the experimental width of (5), we see
that the latter provides a very strong evidence of $\sim 60 \sigma$
for the existence of a top quark as isospin partner of $b$.

It may be noted here that the first three evidences are also based on
the $T^3_b$ dependence of the $Z\bar bb$ coupling.  But they are all
low energy processes, probing the effects of $Z\bar bb$ coupling far
from the $Z$ mass shell.  Thus the effects could be mimicked e.g. by
other gauge bosons $(Z')$ occurring in some extensions of the standard
model gauge group [8].  There is no such ambiguity, however, for the
on-shell $Z\bar bb$ coupling as measured by the decay width (5).

\section{Indirect Estimate of the Top Quark Mass}

There are several indirect constraints on the top quark mass.  The
experimental value of the $B_d - \bar B_d$ mixing [6], which is
dominated by the top quark exchange box diagram, gives a lower mass
limit [10]
$$
m_t \gsim 60 ~{\rm GeV}.
\eqno (9)
$$
A similar lower limit is also obtained from an indirect estimate of
the $W$ boson width at the Tevatron collider [7,11],
$$
\Gamma_W = 2.06 \pm .06 \pm .06 ~{\rm GeV},
\eqno (10)
$$
which is completely saturated by the lighter fermion contributions,
i.e.
$$
\Gamma_W = {\atop {\underbrace{G_F M^3_W \over {6\sqrt{2}\pi}} \atop
{0.23 ~{\rm
GeV}}}} \left[e\bar\nu + \mu\bar\nu + \tau\bar\nu + 3\left(1 + {\alpha_s
\over \pi}\right) (d\bar u + s\bar c)\right].
\eqno (11)
$$
More importantly there is an upper limit on top quark mass,
$$
m_t \lsim 200 ~{\rm GeV},
\eqno (12)
$$
coming from the radiative corrections to $W$ and $Z$ masses [12].
More over the precission measurement of these radiative correction
effects at LEP has recently sharpened this constraint into an indirect
estimate of $m_t$ [13].  Therefore we shall discuss this result in
some detail.

The $W$ mass can be easily calculated at the tree level from the muon
decay diagram of Fig. 1c, and the corresponding $Z$ mass from the tree
level relation\footnote{This relation shall continue to be used as a
working definition of $\sin^2\theta_W$ in the presence of radiative
corrections.}
$$
M_Z = M_W/\cos\theta_W.
\eqno (13)
$$
The observed rate of muon decay gives a precise estimate of the Fermi
coupling
$$
G_F = {g^2 \sqrt{2} \over 8M^2_W} = {\pi \alpha \over \sqrt{2} \sin^2
\theta_W M^2_W},
\eqno (14)
$$
quoted in (7).  The mixing angle, as estimated from the neutrino
scattering experiments [7], is
$$
\sin^2\theta_W = .226 \pm .005.
\eqno (15)
$$
Finally the EM coupling $\alpha$ at the appropriate mass scale is
$$
\alpha (M^2_Z) = \alpha (m^2_e) [1 + \Delta r] \simeq 1/128,
\eqno (16)
$$
corresponding to a EM radiative correction of $\Delta r \simeq
.07$.\footnote{To be more precise $\alpha (M^2_Z) \simeq
1/128.8$ [13], corresponding to a EM radiative correction of $\sim
6$\%.  But there is a small radiative correction of $\sim 1$\% coming
from weak processes other than the top quark exchange term $(\Delta
r')$ discussed below [14].}  From (7) and (13-16) one gets,
$$
M_W = 81.2 \pm 0.8 ~{\rm GeV}, \ \ \ M_Z = 92.2 \pm 0.9 ~{\rm GeV}
\eqno (17)
$$
compared to the experimental values [7] of
$$
M_W = 80.1 \pm 0.4 ~{\rm GeV}, \ \ \ M_Z = 91.19 \pm 0.01 ~{\rm GeV}.
\eqno (18)
$$
The 1\% uncertainty in the tree level predictions of $M_W$ and $M_Z$
are simply reflections of the 2\% uncertainty in the $\sin^2\theta_W$
input (15).  The predictions are higher than the corresponding
experimental values by $1\sigma$.  Consider now the weak radiative
corrections to $M_W$ and $M_Z$.  For large $m_t$, the dominant
corrections come from the $t\bar b$ loop contribution to the $W$
propagator (Fig. 2) and the analogous $t\bar t$ contribution to the $Z$.
The reason is the large Yukawa couplings of longitudinal $W$ and $Z$
bosons to top quark, which are proportional to $m_t$ as remarked
earlier.  The resulting radiative correction is $\propto m^2_t$, i.e.
$$
M^2_W = {\pi \alpha (M^2_Z) \over \sqrt{2} \sin^2 \theta_W G_F} (1 +
\Delta r'),
\eqno (19)
$$
$$
\Delta r' \simeq {-3\sqrt{2} \over 16\pi^2} G_F \cot^2 \theta_W m^2_t
\simeq -1.07 \times 10^{-6} m^2_t.
\eqno (20)
$$
Thus for $m_t > 200 ~{\rm GeV}$ one gets
$$
\Delta r' < -.043 \Rightarrow M_W < 79.5 \pm 0.8 ~{\rm GeV}, \ M_Z <
90.2 \pm 0.9 ~{\rm GeV},
\eqno (21)
$$
i.e. a large negative radiative correction that pushes down $M_Z$
below the experimental value by at least $1\sigma$.  Consequently one
gets the upper limit of (12).

Fig. 3 shows a more exact calculation of the radiative correction as a
function of $m_t$ for two extreme values of the Higgs mass [14].  It
has a mild (logarithmic) dependence on the latter.  The $1\sigma$
bounds obtained from the experimental values of $M_Z$ and
$\sin^2\theta_W$ are shown separately for the cases where
$\sin^2\theta_W$ is estimated from neutrino scattering (15) or from
$M_W/M_Z$ (18).  In either case the favoured value of the radiative
correction is $\Delta r' \simeq -.03$ corresponding to $m_t \simeq 170
{}~{\rm GeV}$.  But the size of the error bar, which reflects the
uncertainty in the estimate of $\sin^2\theta_W$, is too large to give
a precise estimate of $m_t$.  Recently it has been possible to pin
down $m_t$ or equivalently the $\sin^2\theta_W$ more precisely from a
global fit to the precission measurements of the $Z$ parameters at
LEP [13].  It gives
$$
m_t = 173^{+12 \ \ \ +18}_{-13 \ \ \ -20} {\rm GeV},
\eqno (22)
$$
or equivalently
$$
\sin^2\theta_W = .2249 \pm .0013^{+.0003}_{-.0002},
\eqno (23)
$$
where the second errors correspond to varying $M_H$ from 1000 GeV
(upper) to 60 GeV (lower).  This result is shown as a hatched band in
Fig. 3.  Note that the width of this band, or equivalently the 1st
error bar of (22), corresponds to the uncertainty in the estimate of
$m_t$ related to that of $\sin^2 \theta_W$.  This uncertainty is now
very small; but there is a somewhat larger uncertainty coming from the
unknown Higgs mass.  Consequently the overall uncertainty in the
indirect estimate of $m_t$ from LEP is somewhat larger than the direct
estimate from the CDF experiment; but there is remarkably good
agreement between the two results.

\section{Direct Top Quark Search at the Electron-positron Collider}

The $e^+e^-$ collider can
provide the cleanest signal for top quark; but unfortunately the energies
are too low.  The simplest way to look for $e^+e^- \rightarrow \bar t t$
(Fig. 1b) is through the ratio of cross-sections
$$
R = {\sigma(e^+e^- \rightarrow {\rm hadrons}) \over \sigma(e^+e^-
\rightarrow \mu^+\mu^-)} = {\sigma(e^+e^- \rightarrow \Sigma \bar q q)
\over \sigma(e^+e^- \rightarrow \mu^+ \mu^-)} \simeq 3 \sum Q^2_q,
\eqno (24)
$$
which should show a jump of $\Delta R = 3Q^2_t = 4/3$ units across the
$\bar t t$ threshold.  One can readily check that this corresponds to
an increase of the hadronic cross-section or $R$ by about one third.
The second way is to look at the event shape.  The
lighter quark pairs fly off back to back carrying the total centre of mass
energy and thus give rise to highly collinear events.  In contrast, near
the $t \bar t$ threshold, the heavy $t$ quark pair will be produced
practically at rest; and each will decay into 3 quarks (Fig. 1c)
$$
t \rightarrow bu\bar d, bc\bar s.
\eqno (25)
$$
Thus the total centre of mass energy would be shared amongst the 6 light
quarks, giving rise to a more spherical (isotropic) event.

The PETRA and TRISTAN colliders have looked for $e^+e^- \rightarrow \bar t
t$ events using these methods and found none.  Thus they give lower bounds
on top quark mass equal to their respective beam energies.  The larger
one, coming from TRISTAN [15] is $m_t > 26$ GeV.  More recently the
LEP $e^+e^-$ collider has increased this mass bound to [16]
$$
m_t > 45~{\rm GeV},
\eqno (26)
$$
which corresponds to its beam energy.  With the LEP-II, scheduled for
the late nineties, the probe can be further extended upto its beam
energy of about 90 GeV.  As we have already seen, however, this is
not large enough.

\section{Direct Top Quark Signal at the Antiproton-proton Collider}

The $\bar p p$ collider is best suited for a heavy top quark search
because of its higher energy reach.  But the signal is messy; and one
has to use special techniques to
disentangle it from the background.  The dominant mechanism for top quark
production is the so-called flavour creation process of gluon-gluon fusion
(Fig. 4) and quark-antiquark fusion (Fig. 1a), i.e.
$$
gg (\bar q q) \rightarrow \bar t t.
\eqno (27)
$$
The best way to look for top is to look for a prompt charged lepton
$\ell$ (i.e. $e$ or $\mu$) coming from its leptonic decay
$$
t \rightarrow b\nu\ell
\eqno (28)
$$
as per Fig. 1c.  This eliminates the background from gluon and
ordinary stable quarks $(u,d,s)$. Of course the charged lepton could
come from the unstable quarks $b$ and $c$ i.e.
$$
gg(\bar qq) \rightarrow \bar bb,\bar cc;
$$
$$
b \rightarrow c\nu\ell, \ c \rightarrow s\nu\ell.
\eqno (29)
$$
These background can be effectively suppressed by requiring the
charged lepton to be isolated from the other particles.  Because of
the large energy release in the decay of the massive top quark, the
decay products come wide apart.  In contrast the energy release in the
light $b$ or $c$ quark decay is small, so that the decay products come
together in a narrow cone -- i.e. the charged lepton appears as a part
of the decay quark jet.  The isolated lepton signature provides a
simple but very effective signature for top quark, first suggested in
[17].  Using this signature the top quark search was carried out at
the CERN $\bar pp$ collider and then at the Tevatron collider upto a
mass limit of $m_t \simeq 90$ GeV [18].

With the luminosity upgrade of the Tevatron collider it has been
possible now to extend the search to the mass range of $100 - 200$
GeV.  A top quark in this mass range decays into a real $W$ boson, so
that one has a $2W$ final state, i.e.
$$
t\bar t \rightarrow W^+W^-b\bar b.
\eqno (30)
$$
The resulting signature for top quark and the corresponding background
were first analysed in [19].  Requiring leptonic decay of both or one
of the $W$ bosons leads to an isolated dilepton or single lepton
signature, i.e.
$$
t\bar t \rightarrow \ell^+ \ell^-  \nu\bar \nu  b\bar b
\eqno (31)
$$
or
$$
t\bar t \rightarrow \ell\nu b\bar b q{\bar q}'.
\eqno (32)
$$
In either case there are several accompanying quark jets and a
missing-$p_T$ due to the escaping neutrino(s).  The dilepton signature
(31) is small in size, since one has to pay the price of a small
leptonic branching ratio of $W$ $(\simeq 2/9)$ twice.  But it is
relatively clean, since there is only a small background from the
second order electroweak process
$$
q\bar q \rightarrow W^+W^-.
\eqno (33)
$$
In contrast the single lepton signature (32) is relatively large; but
one has to contend with a much larger background from single $W$
production along with QCD jets, e.g.
$$
\bar qq \rightarrow Wgg \rightarrow \ell\nu j_1 j_2.
\eqno (34)
$$
However the QCD jets are normally soft and besides one has to pay a
price of $\alpha_s$ for each additional jet.  In contrast the decay of
a heavy $t\bar t$ pair automatically gives a large number of decay
quark jets in (32), which are hard and well separated in angle.
Consequently the background can be kept in control by suitable cuts on
the number and hardness of the jets accompanying the isolated lepton
[19-21].

Fig. 5 shows the predicted dilepton and single lepton signals from
[19] for an integrated Tevatron collider luminosity of $100 ~{\rm
pb}^{-1}$, which is relevant for its current run [22,23].  The
dilepton signal is seen to be viable upto $m_t \simeq 150~{\rm GeV}$.
The single lepton signal (32) is shown separately for different
numbers $(n = 2,3,4)$ of accompanying jets.\footnote{This separation
depends to some extent on the choice of cone angle and $p_T$ threshold
of the jets.  The cone angle used by the Tevatron experiments [22,23]
are some what smaller than the conservative assumption of [19].  As a
result the 3 and 4 jet contributions become comparable for $m_t = 150
- 200$ GeV.}  The single lepton background from (33) and (34) are also
shown for comparison.  Modest jet hardness cuts of
$$
\sum \vec p_{Tj} \ > \ 60 ~{\rm GeV}, \ \ \ m_{jj} \ > \ 60 ~{\rm
GeV},
\eqno (35)
$$
have been applied on the vector sum of the jet $p_T$'s and the
invariant mass of the two hardest jets.  This is adequate to keep the
background below the level of the $n \geq 2$ signal upto $m_t \simeq
150$ GeV.  Moreover it is possible to achieve this all the way upto
$m_t \simeq 200$ GeV using a tighter jet hardness $(\displaystyle \sum
\vec p_{Tj})$ cut and restricting to $n \geq 3$, as noted in [19].
Similar results have been obtained in [20,21] using alternative forms
of the jet hardness variable like the scalar sum of the jet $p_T$'s
[20] or the $p_T$'s of the two hardest jets [21].  Finally, the
presence of a pair of $b$ jets in the signal process can be used to
separate it from the background, given a good $b$ identification
efficiency via a microvertex detector.

Recently the CDF and D${\rm O}\!\!\!\!/$ experiments, working at the
Tevatron collider, have identified a heavy top quark signal using the
above mentioned techniques [22,23].  Their results are based on the
integrated luminosities of 67 and 50 pb$^{-1}$ respectively, i.e.
about half the projected luminosity of $\sim 100$ pb$^{-1}$ for the
current Tevatron run.  The CDF experiment [22] uses $b$ tagging via a
silicon microvertex detector to separate the single lepton signal (32)
from the QCD background (34), while the D${\rm O}\!\!\!\!/$ experiment
[23] achives this using the jet hardness criterion.  Fig. 6 shows the
CDF data for the $W$ plus $\geq \ 4$ jet events before $b$-tagging
along with the predicted background [22].  The events are plotted
against the reconstructed mass, i.e. the invariant mass of the $W$ and
a suitably chosen jet so that it nearly matches with the invariant
mass of the three remaining jets.  Although the background accounts
for about 70\% of the data, one sees a clear excess in the mass range
of $\sim 175$ GeV.  Fig. 8 shows the corresponding events after
$b$-tagging.  The later improves the signal to background ratio
significantly, while the signal size is reduced by the tagging
efficiency factor of $\sim 40$\%.  Thus a top quark signal of about a
dozen events is clearly visible against a background of $\sim 7$
events.  The shape of the event distribution gives a top quark mass of
$$
m_t = 176 \pm 8 \pm 10 \ {\rm GeV},
\eqno (36)
$$
where the 1st and 2nd errors correspond to the statistical and
systematic uncertainties respectively.  The $t\bar t$ cross-section
estimated from the signal size,
$$
\sigma_{t\bar t} = 6.8^{+3.6}_{-2.4} \ {\rm pb},
\eqno (37)
$$
is in agreement with the QCD prediction.  The statistical significance
of the CDF top quark signal after combining the single lepton and
dilepton data is at the level of $4.8\sigma$.  The D${\rm O}\!\!\!\!/$
experiment has given a top quark signal of comparable statistical
significance and similar size of cross-section; but their mass
estimate is less precise [23]; i.e.
$$
m_t = 199^{+19}_{-21} \pm 22 \ {\rm GeV}.
\eqno (38)
$$

Let me try to put the above result in perspective and make some future
projections.  As we have seen in the earlier sections, the top quark
search has been a long programme extending over the last fifteen
years.  The evidence for the existence and mass of the top quark has
built up step by step from a large number of indirect experiments.
The direct observation of the top quark signal at the Tevatron
collider is of course the final step.  It is also the most difficult
one.  Firstly we see from Fig. 5 that the relevant signal size for
$m_t \sim 175$ GeV is only a few tenths of a $pb$.  Compared with the
$\bar pp$ total cross-section of $\sim 100$ mb, this is at the level
of a few parts in a trillion.  This is a thousand time smaller than
the $W$ and $Z$ boson signals, observed a decade ago, which were a few
parts in a billion.  Secondly, the $W$ and $Z$ bosons had unmistakable
leptonic signatures with practically no background.  In contrast, the
direct $W$ production is an unavoidable background for the top quark
signal, which is hard to suppress and impossible to eliminate.  Thus
one has to extricate the signal events of few parts in a trillion from
the background of few parts in a billion.  Finally it is the first
example of identifying a new particle peak in a multijet in stead of a
multiparticle channel.  It may be noted here that the $W$ and $Z$
boson peaks are yet to be seen in a dijet channel at a hadron
collider.  One should bear these points in mind in order to appreciate
the real significance of the top quark signal, observed at Tevatron.

At the same time it is fair to say that the current Tevatron result is a
semifinal rather than the final step in the top quark search.  For a
$5\sigma$ signal by itself constitutes a promising rather than
conclusive signal.  The recent history of particle physics is replete
with many examples of $5\sigma$ signals that have fallen by the way
side.  The reason the top quark signal is taken more seriously of
course is that it falls in place with the indirect evidences,
particularly from LEP (Fig. 3).  Nonetheless there is a lot of scope
for improvement and cross-checks on the direct top quark signal, which
can be done with more data from the Tevatron collider.  For instance,
one can check the correlation between the event samples selected via
$b$-tagging and via the jet hardness cut.  Moreover one can supplement
$b$-tagging by a jet hardness cut to improve upon the signal to
background ratio shown in Fig. 7, at a cost to the signal size of
course.  Besides one should be able to get independent signals at $> \
5\sigma$ levels in single lepton and dilepton (particularly $e\mu$)
channels, and check their relative magnitudes with the universality
prediction.  Some of these can surely be done with the doubling of the
data sample by the end of the current run.  Note however that one
would still have only about two dozens of signal events.  On the other
hand the installation of the main injector following this run is
expected boost the Tevatron luminosity further by an order of
magnitude.  Consequently the next run, scheduled for 1998, is expected
to yield a data sample of a few hundred signal events.  This will
enable one to improve the signal to background ratio and perform
various cross-checks, as indicated above.  As a result one expects to
see a conclusive signal for top from this data.  Of course there will
still be a lot of interesting top quark physics left for LHC and
beyond.

\section{Top Quark Physics at LHC and NLC}

The top quark production cross-section at LHC is $\sim 100$ times
larger than at the Tevatron collider energy.  Fig. 8 shows the
expected $t\bar t$ cross-section in the cleanest dilepton channel
$(e\mu)$ against the $p_T$ of the softer lepton [24].  It corresponds
to an integrated cross-section of $\sim 10^4$ fb in the $e\mu$ channel
or equivalently $\sim 10^5$ fb in the single-lepton channel discussed
above.  Even with the low luminosity option of LHC ($\sim 10$
fb/year), this would imply an annual rate of $\sim 1$ million top
quark events -- i.e. similar to the rate of $Z$ events at LEP.  Thus
the LHC can serve as a top quark factory.  This will enable one to
study its decay properties in detail and to search for new particles
in the top quark decay.  In particular there has been a good deal of
recent interest in the search for one such new particle, for which the
top quark decay offers by far the best discovery limit -- i.e. the
charged Higgs boson $H^\pm$ of the supersymmetric standard model.
Detailed signatures for $H^\pm$ search in the top quark decay at LHC
have been studied in [25].

Of course the ultimate stage of the top quark physics will be reached
at the next linear collider (NLC), a generic name for a $e^+e^-$
machine with CM energy $\gsim 500$ GeV, which is hoped to follow LHC
[26].  As we have already seen in eq. (24) above, the $t\bar t$
production cross-section would constitute about a quarter of the
hadronic cross-section in a $e^+e^-$ collider.  The copious production
rate and the clean environment of a $e^+e^-$ machine would make it
possible to measure top quark mass to an accuracy of 0.5 GeV and
measure its life time.  It may be noted here that the large mass of
top implies a life time $\sim 10^{-23}$ sec.; which means that the top
quark decays before hadronization.  Consequently the spin and
polarisation of top can be measured from the kinematic distribution of
its decay products.  The polarisation information will be useful for
studying CP violation effects in $t\bar t$ production.  In particular
this process is well suited to look for possible CP violation induced
via the Higgs sector because of the large Higgs coupling to top.  Of
course the short life of top means there will be no toponium states to
study.  Nonetheless there is a wealth of information to be gained from
the detailed study of $t\bar t$ production at the NLC.

\newpage

\begin{center}
\underbar{\bf References}
\end{center}
\bigskip
\begin{enumerate}
\item[{1.}] See any text on particle physics, e.g. F. Halen and A.D.
Martin, Quarks and Lepton, John Wiley, New York (1984).
\item[{2.}] S.L. Wu, Proc. of 1987 Intl. Symp. on Lepton and Photon
Int. at High Energies, North-Holland (1988).
\item[{3.}] V. Barger and S. Pakvasa, Phys. Lett. 81B, 195 (1979); \\
G.L. Kane and M.E. Peskin, Nucl. Phys. B195, 29 (1982).
\item[{4.}] CLEO collaboration: A. Bean et al., Phys. Rev. D35, 3533
(1987).
\item[{5.}] D.P. Roy and S. Uma Sankar, Phys. Lett. 243B, 296 (1990).
\item[{6.}] ARGUS collaboration: H. Albercht et al., Phys. Lett. 192B,
245 (1987); \\
CLEO collaboration: M. Artuso et al., Phys. Rev. Lett. 62, 2233
(1989).
\item[{7.}] Review of Particle Properties, Phys. Rev. D50, 1173-1826
(1994).
\item[{8.}] S. Pakvasa, D.P. Roy and S. Uma Sankar, Phys. Rev. D42,
3106 (1990)
\item[{9.}] D.P. Roy, in Particle Phenomenology in the 90's (Proc. of
WHEPP-II, Calcutta, 1991), World Scientific (1992), p 30.
\item[{10.}] P.J. Franzini, Phys. Rep. 173, 1 (1989), and references
therein.
\item[{11.}] CDF collaboration: F. Abe et al., Phys. Rev. Lett. 73,
220 (1994).
\item[{12.}] U. Amaldi et al., Phys. Rev. D36, 1385 (1987); \\
G. Costa et al., Nucl. Phys. B297, 244 (1988); \\
J. Ellis and G. Fogli, Phys. Lett. 232B, 139 (1989); \\
P. Langacker, Phys. Rev. Lett. 63, 1920 (1989).
\item[{13.}] D. Schaile, Proc. of 27th Intl. Conf. on High Energy
Physics, Glasgow, 1994, Vol. I, p 27 (Inst. of Physics Publishing,
Bristol).
\item[{14.}] W. Hollik, Karlsruhe Preprint, KA-TP-2-1995 (1995).
\item[{15.}] VENUS collaboration: H. Yoshida et al., Phys. Lett. 198B,
570 (1987); \\ TOPAZ collaboration: J. Adachi et al., Phys. Rev. Lett.
60, 97 (1988); \\ AMY collaboration: H. Sagawa et al., Phys. Rev.
Lett. 60, 93 (1988).
\item[{16.}] OPAL collaboration: M. Akrawy et al., Phys. Lett. 236B,
364 (1990); \\ ALEPH collaboration: D. Decamp et al., Phys. Lett.
236B, 511 (1990); \\ DELPHI collaboration: P. Abreu et al., Phys.
Lett. 242B, 536 (1990).
\item[{17.}] R.M. Godbole, S. Pakvasa and D.P. Roy, Phys. Rev. Lett.
50, 1539 (1983); \\ V. Barger, A.D. Martin and R.J.N. Phillips, Phys.
Rev. D28, 145 (1983).
\item[{18.}] CDF collaboration: F. Abe et al., Phys. Rev. D45, 3921
(1992).
\item[{19.}] S. Gupta and D.P. Roy, Z. Phys. C39, 417 (1988).
\item[{20.}] H. Baer, V. Barger and R.J.N. Phillips, Phys. Rev. D39,
3310 (1989).
\item[{21.}] R. Wagner et al., Top quark working group report in
Physics at the Fermilab in the 1990's (Proc. of the Breckenbridge
Workshop), World Scientific (1990), p 181.
\item[{22.}] CDF collaboration: F. Abe et al., Phys. Rev. Lett. 74,
2626 (1995).
\item[{23.}] D${\rm O}\!\!\!\!/$ collaboration: S. Abachi et al.,
Phys. Rev. Lett. 74, 2632 (1995).
\item[{24.}] N.K. Mondal and D.P. Roy, Phys. Rev. D49, 183 (1994).
\item[{25.}] D.P. Roy, Phys. Lett. 277B, 183 (1992); 283B, 403 (1992).
\item[{26.}] See e.g. Proc. of the Workshop ``$e^+e^-$ collisions at
500 GeV: The Physics Potential'', DESY Pub. 92-123A/B (1992).
\end{enumerate}
\newpage

\begin{enumerate}
\item[{Fig. 1.}] Basic interactions of quarks and leptons.  a, Strong.
b, Electromagnetic. c, Charged current weak. d, Neutral current weak.
\item[{Fig. 2.}] Radiative correction to the $W$ boson mass arising
from the $t\bar b$ loop.
\item[{Fig. 3.}] Radiative correction as a function of top mass for
$M_H = 60$ and $1000$ GeV.  The $1\sigma$ bound from $M_Z$ and
$\sin^2\theta_W$ are shown for three different estimates of
$\sin^2\theta_W$: from $M_W/M_Z$ (horizontal band), from $\nu N$
scattering (point) and finally from the precission measurement of $Z$
parameters at LEP (hatched band) [13,14].  The direct estimate of top
mass from the CDF experiment [22] is also shown for comparison.
\item[{Fig. 4.}] Top quark production in $\bar pp$ ($pp$) collision
via gluon-gluon fusion.
\item[{Fig. 5.}] Top quark contribution to the isolated lepton plus
$n$-jet events and also dilepton events (dotted line) shown for the
typical energy (2 TeV) and luminosity (100 pb$^{-1}$) of the Tevatron
collider.  The background to the 2-jet events from $W$ plus 2-jet and
$W$ pair production processes are also shown. [19]
\item[{Fig. 6.}] Reconstructed mass distribution for the $W + \geq$
4-jet sample prior to $b$-tagging (solid).  Also shown is the
background distribution (shaded), with the normalization constrained
to the calculated value. [22]
\item[{Fig. 7.}] Reconstructed mass distribution for the $b$-tagged $W
+ \geq$ 4-jet events (solid).  Also shown are the background shape
(dotted) and the sum of background plus $t\bar t$ Monte Carlo for
$M_{top} = 175$ GeV/$c^2$ (dashed), with the background constrained to
the calculated value, $6.9^{+2.5}_{-1.9}$ events.  The inset shows the
likelihood fit used to determine the top mass. [22]
\item[{Fig. 8.}] The expected $t\bar t$ signal at LHC in the cleanest
$(e\mu)$ channel shown against the $p_T$ of the 2nd (softer) lepton.
The $b\bar b$ background with and without the isolation cut are also
shown. [24]
\end{enumerate}

\end{document}